\documentclass[]{article}
\usepackage{graphicx}
\usepackage{natbib}
\usepackage{authblk}
\usepackage[a4paper]{geometry}

\begin{document}
\title{Gut microbiome composition: back to baseline?}
\author[1$\dagger$]{Matthias M. Fischer}
\author[1]{Matthias Bild}
\affil[1]{\small Freie Universit\"at Berlin; Fachbereich Biologie, Chemie, Pharmazie; Institut f\"ur Biologie, Mikrobiologie; 14195 Berlin, Germany}
\affil[$\dagger$]{To whom correspondence should be addressed. Tel.: +49 30 838 53373, E-Mail: \texttt{m.m.fischer@fu-berlin.de}}
\date{}
\maketitle

\begin{abstract}
In Nature Microbiology, \textit{Palleja} and colleagues studied the changes in gut microbiome composition in twelve healthy men over a period of six months following an antibiotic intervention. The authors argued that the \textit{gut microbiota of the subjects recovered to near-baseline composition within 1.5 months} and only exhibited a \textit{mild yet long-lasting imprint following antibiotics exposure}. We here present a series of re-analyses of their original data which demonstrate a significant loss of microbial taxa even after the complete study period of 180 days. Additionally we show that the composition of the microbiomes after the complete study period only moderately correlates with the initial baseline states. Taken together with the lack of significant compositional differences between day 42 and day 180, we think that these findings suggest the convergence of the microbiomes to another stable composition, which is different from the pre-treatment states, instead of a recovery of the baseline state. Given the accumulating evidence of the role of microbiome perturbations in a variety of infectious and non-infectious diseases, as well as the crucial role antibiotics play in modern medicine, we consider these differences in compositional states worthy of further investigation. \\
\end{abstract}

In Nature Microbiology, \textit{Palleja} and colleagues studied the changes in gut microbiome composition in twelve healthy men over a period of six months following a four-day intervention with a cocktail of three last-resort antibiotics by using shotgun sequencing-based metagenomics \citep{ori}. The authors argued that the \textit{gut microbiota of the subjects recovered to near-baseline composition within 1.5 months.} Subsequently, they claimed that \textit{the gut microbiota of healthy young adults are resilient to a short-term broad-spectrum antibiotics intervention} despite a \textit{mild yet long-lasting imprint following antibiotics exposure}. \\

We do appreciate the work of the authors and agree that studying the effects of antimicrobials on the composition and function of the human microbiome is promising from a medical point of view. Additionally, we agree with the authors that the gut microbiome of healthy young adults seems to possess a substantial amount of resilience towards disturbances such as the exposure to antibiotic substances. However, we are sceptical about the specific interpretation of their data implying a return of the examined bacterial communities to a state of near-baseline composition. \\

The composition of the gut microbiome is characterised by a high degree of variation between individuals \citep{individual}, which makes a within-subject analysis of the experimental data preferable due to a higher statistical power \citep{within}. Raw species counts of the examined $n=12$ patients drastically decreased from $160 \pm 5.9$ (SE) species before the intervention to $73 \pm 3.2$ species eight days after the intervention. After 180 days, the raw species counts amounted to $136 \pm 9.1$. A pairwise comparison of species counts between day zero and day 180 revealed a significantly reduced species count with an average difference of 24 species (paired Wilcoxon Rank Sum Test, $V=70$, $p<0.05$*). On average, this difference was equivalent to a relative reduction of the species pool size by $(15 \pm 5.0) \; \%$. \\

For each of the patients, we additionally performed a Bray-Curtis ordination of the metagenomic data and assessed the dissimilarity of the samples compared to the baseline over time (Figure 1). We found the highest dissimilarity immediately after treatment with the average Bray-Curtis distance on day eight rising to $0.93 \pm 0.014$. After the complete 180 days, the average Bray-Curtis distance still amounted to $0.64 \pm 0.044$, indicating significant differences from the baseline composition. A similar picture emerges, when Kendall's rank correlation coefficient $\tau$ is used as similarity metric of the relative species abundances instead. On day eight, the average correlation coefficient has fallen to $0.25 \pm 0.032$; at the end of the complete 180 days, the average $\tau$ of $0.56 \pm 0.022$ indicates only a moderately high correlation with the baseline abundance structure. Figure 2 shows correlation plots of relative microbial abundances on day 180 compared to day zero for each individual patient. \\

\begin{figure}[!h]
	\centering
	\includegraphics[width=8cm]{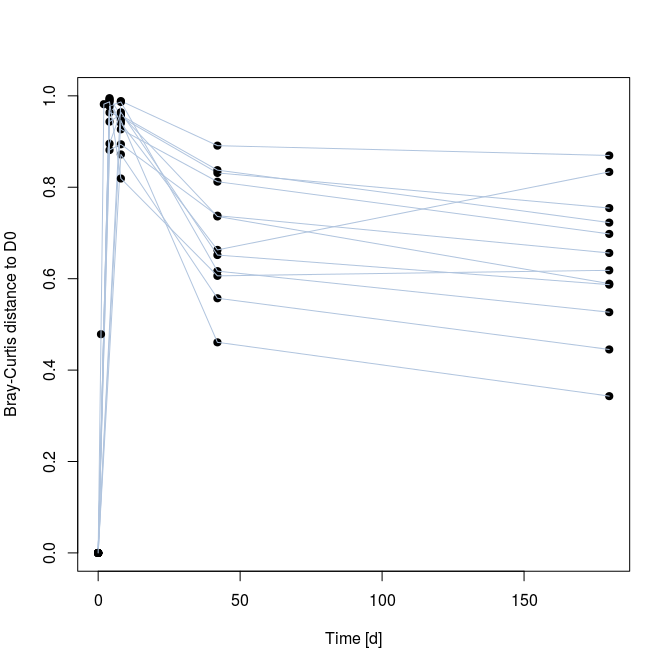}
	\caption{\textit{Bray-Curtis distances between the gut microbiomes of the twelve patients compared to the baseline composition at day zero.}}
\end{figure}

\begin{figure}[!h]
	\centering
	\includegraphics[width=15cm]{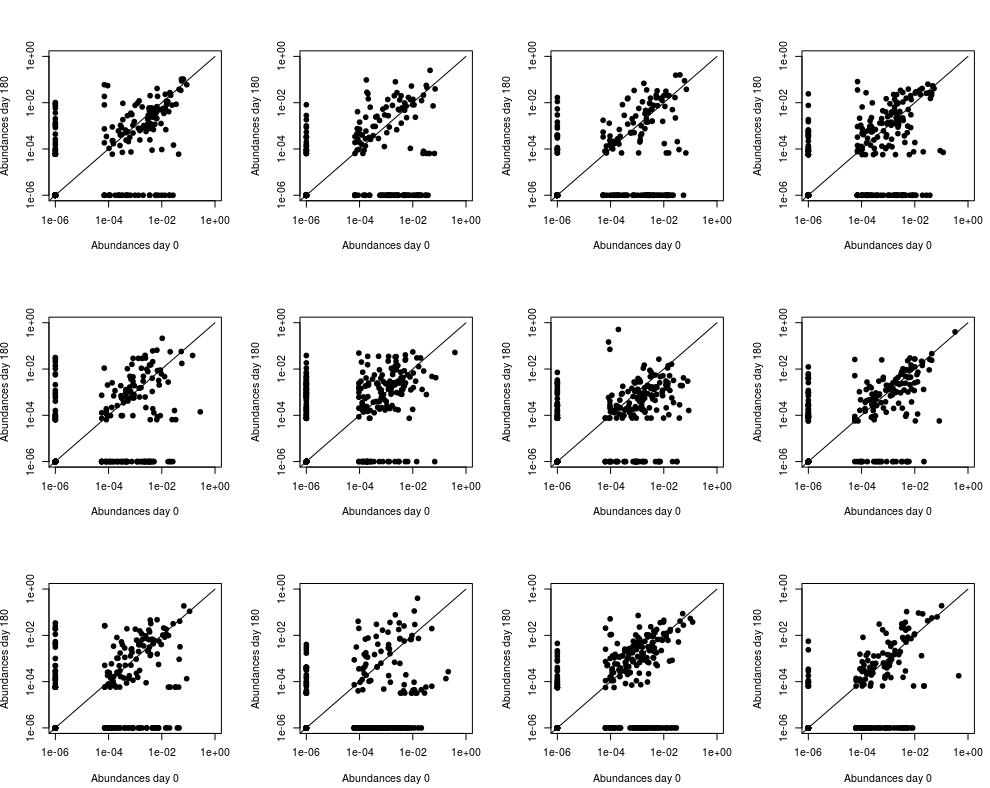}
	\caption{\textit{Correlation plots of relative microbial abundances on day 180 (y-axis) compared to day zero (x-axis) for each individual patient. Note the logarithmic scale of both axes. Abundances of zero are shown as $10^{-6}$.}}
\end{figure}

Overall, we believe the experimental approach of \textit{Palleja} and colleagues to be valid and suitable, and the examined problems important and worthy of investigation. However, given the significantly reduced raw species counts even after 180 days, as well as the merely moderate correlation of the final abundance structures compared to day zero, we do not agree that their data convincingly support the claim that the gut microbiomes of the examined patients have converged to a near-baseline composition within only six weeks after the intervention. Unfortunately, the authors have not specified what they consider 'near', however, using the statistical approach in this comment, we demonstrated significant differences, even given the limited sample size. In fact, the lack of significant compositional differences between day 42 and day 180 which the authors have shown even suggests the convergence of the microbiome to another stable composition completely different from the baseline state. Of course the important question of the biological and practical significance of such differences compared to the baseline state remains open. \\

Additionally, we would like to propose extending future similar studies to longer time spans in order to also understand the long-term recovery dynamics of the gut microbiome towards external stressors in more detail. A recent study by \textit{Shaw} and colleagues has argued that the microbiome can be conceptualised as sitting in a 'stability landscape' with multiple stable equilibria, and that sufficiently strong perturbations are able to shift the microbiome from its normal equilibrium state to another one \citep{Shaw}. It remains an interesting open question whether the 'stable composition' identified by \textit{Palleja} and colleagues is indeed such an alternative stable state or will instead gradually converge back to the normal equilibrium state over time. For answering this question, it might also be a worthwhile approach to sample individual patients not only once, but instead repeatedly at each time point in order to reliably quantify intra-patient variability as well.

\paragraph{Author contributions}
MMF analysed the data and wrote the manuscript. MB provided critical feedback and additions to an earlier version of this work. Both authors have read the final version of the manuscript and approve of it.

\paragraph{Competing interests}
Both authors declare no competing interests.

\bibliographystyle{apalike}
\bibliography{references}

\end{document}